\documentstyle[prl,floats,aps]{revtex}
\input psfig
\pssilent
\begin{document}
\input psfig \pssilent 
\title{Consistent canonical quantization of general relativity in the 
space of Vassiliev knot invariants}
\author{
Cayetano Di Bartolo$^{1}$
Rodolfo Gambini$^{2}$, Jorge Griego$^2$, Jorge Pullin$^3$}
\address{1. Departamento de F\'{\i}sica, Universidad Sim\'on Bol\'{\i}var,
Aptdo. 89000, Caracas 1080-A, Venezuela.}
\address{2. Instituto de F\'{\i}sica, Facultad de Ciencias, 
Igu\'a 4225, esq. Mataojo, Montevideo, Uruguay.}
\address{ 3. Center for Gravitational Physics and Geometry, 
Department of Physics,\\ 
The Pennsylvania State University, 104 Davey Lab, University Park, PA
16802.} 
\date{September 17th 1999}
\maketitle
\begin{abstract}
We present a quantization of the Hamiltonian and diffeomorphism constraint
of canonical quantum gravity in the spin network representation. The novelty
consists in considering a space of wavefunctions based on the Vassiliev
knot invariants. The constraints are finite, well defined, and reproduce
at the level of quantum commutators the Poisson algebra of constraints of 
the classical theory. A similar construction can be carried out in
$2+1$  dimensions leading to the correct quantum theory.
\end{abstract}

The Ashtekar new variables \cite{As86} 
describe general relativity as a theory of a connection, having the
same kinematical phase space as a Yang--Mills theory. The canonical
conjugate pair is given by a set of (densitized) triads
$\tilde{E}^a_i$ and an SU(2) connection $A_a^i$. This allowed to 
describe the theory in terms of holonomies \cite{RoSm90} leading to the
development of the loop representation, and later the spin network
representation \cite{RoSmspin}. These representations encode
in a natural way the diffeomorphism invariance of the theory  through
the notion of knot invariance. The dynamics of the theory, embodied in
the Hamiltonian constraint, remained elusive. The quantization of this
constraint led to the so-called Wheeler-DeWitt equation in the
traditional formulation of general relativity. This is a
non-polynomial equation and presents several challenges as a quantum
field theory, since the usual techniques for regularizing operators
introduce fiducial background metric structures that are incompatible
with the general covariance of the theory. In terms of the Ashtekar
new variables an important step forward was realized when Thiemann 
\cite{QSD1}
showed how to write the Hamiltonian constraint as a scalar on
the manifold. This raised hopes that a natural realization in terms of
spin networks could be achieved. Thiemann represented the action of this
constraint  on diffeomorphism invariant states. He showed that the
constraint commuted with itself, as one expects in a diffeomorphism
invariant context. Moreover, Thiemann's formulation took place in the
context of the real version of the Ashtekar variables introduced by
Barbero \cite{Ba}, bypassing the controversial issue of the ``reality
conditions''. The Hamiltonian considered corresponded to the usual
real, Lorentzian general relativity.

In this paper we will present a realization of the Hamiltonian
constraint in terms of a different space of wavefunctions, associated
with the Vassiliev knot invariants. A distinctive feature of these
wavefunctions is that they are ``loop differentiable''. The loop
derivative \cite{GaPubook} is the derivative that arises in the space of
functions of loops
when one considers the change in wavefunctions due to the addition of
an infinitesimal loop. In the context of holonomies, this derivative
encodes the information of the curvature tensor $F_{ab}$. There is a
well known difficulty with computing this derivative in the context of
knot invariants, since due to the diffeomorphism symmetry there is no
notion of ``infinitesimal'' loop. Therefore one cannot compute the
limit involved in the derivative in a direct way. In the case of
Vassiliev invariants one can assign a value to this limit recalling
the relationship between them and the expectation value of the Wilson
loop in a Chern--Simons theory,
\begin{equation}
E(s,\kappa) = \int DA \exp\left( -{1 \over \kappa} {\rm Tr}(A\wedge \partial A+
{2 \over 3} A\wedge A\wedge A)\right) W_s[A]
\end{equation}
where $s$ is a spin network (a multivalent graph with holonomies in
representations of SU(2) associated with each edge) and $W_s[A]$ is an
SU(2) invariant obtained by interconnecting the holonomies along the
edges with appropriate intertwiners constructed with invariant tensors
in the group. It is a natural generalization to the spin network
context of the ``Wilson loop'' (trace of the holonomy) one constructs
with ordinary loops. The quantity $E(s,k)$ is an infinite series in
powers of $1\over k$, and is a (framing dependent) knot
invariant. This invariant was first considered as connected with a
Chern--Simons theory by Witten \cite{Wi88} in the context of loops and
remarkably, also in the context of spin networks \cite{Wi89,Ma}. In
the context of loops this invariant is associated with the evaluation
for a particular value of the variable of the Kauffman polynomial. The
coefficients in the infinite series are all knot invariants and one
can isolate within these coefficients the elements of a basis of
framing independent invariants called the Vassiliev invariants when
restricted to ordinary loops. This
construction can be extended to the spin network context, as we showed
in two recent papers \cite{GaGrPu98,DiGaGrPu99}.  We will refer to the 
resulting invariants as Vassiliev invariants (including the framing
dependent ones), although it should be noticed that this is a generalization 
of the usual notion of Vassiliev invariant, which is customarily
introduced
for ordinary non-intersecting loops. One can evaluate
the loop derivative on these invariants and one is left with a simple
formula \cite{GaGrPu98},
\begin{equation}
\Delta_{ab}(\pi_o^x) E(s,\kappa) =  \kappa 
\sum_{e_k} (-1)^{2(J_j+J_k)} \Lambda_{J_jJ_k}  \epsilon_{abc}
\int_{e_k} dy^c \delta^3(x-y)
E\left(s',\kappa\right).\label{deltae} 
\end{equation}
where $s'$ is a new spin network obtained by interconnecting in a
certain way the original spin network $s$ with the path $\pi$ on which
the loop derivative $\Delta_{ab}$ depends, and $\Lambda_{J_jJ_k}$ is a
group factor dependent on the valences $J_j$ and $J_k$ of the lines
$e_j$ and $e_k$.  The action of the derivative is distributional, as
one would expect in a diffeomorphism invariant context. A similar
action is obtained not just for the infinite series $E$ but also for
each individual coefficient and its framing dependent and framing
independent portions.

In terms of the loop derivative we just discussed one can now obtain
an action for the Hamiltonian constraint in the scalar version introduced by
Thiemann \cite{QSD1}.  We will only discuss for simplicity here the action on
trivalent spin networks and we will concentrate on the ``Euclidean''
portion of the constraint. Thiemann has shown how if one has the
action of this portion one can construct the rest of the full
Lorentzian Hamiltonian constraint. Classically, the constraint is
written as \cite{QSD1},
\begin{equation}
H(N) = {2 \over G} \int d^3x N(x) \{A_a^i,V\} F_{bc}^i \tilde{\epsilon}^{abc},
\end{equation}
where $V$ is the volume of the manifold and $G$ is Newton's
constant. At a quantum level, one introduces a triangulation adapted
to the spin network of the state one is acting upon, replaces the
Poisson bracket by a commutator, and represents the connection as an
infinitesimal holonomy. In the context of trivalent intersections only
one term in the commutator is non-vanishing, and one gets for the
Hamiltonian \cite{DiGaGrPu99},
\begin{equation}
H(N) \psi \left(
\raisebox{-10mm}{\psfig{file=ver3.eps,height=20mm}}\right) = 
{8 \over 3G}
\lim_{\epsilon\rightarrow 0} \int d^3y \sum_{v\in s} 
\epsilon_{ijk} \int_{e_i} du^a \int_{e_j} dw^b 
\chi(u,w,y;v)
N(y) 
\rho(J_1,J_2,J_3)
\Delta^{(k)}_{ab}(\pi_{v}^y)
\psi
\left(
\raisebox{-10mm}{\psfig{file=ver3.eps,height=20mm}}\right), \label{hgen}
\end{equation}
where $\rho$ is a group factor dependent on the valences of the
three incoming lines at the intersection. The action of the Hamiltonian is
only non-vanishing at intersections.  The function $\chi$ is a
regulator that restricts the integrals in $u,w$ to the tetrahedra
surrounding the vertex $v$ and fixes the point $y$ to the vertex
$v$, a concrete realization is 
$\chi(y,z,w)=\Theta_\Delta(y,v)\Theta_\Delta(z,v)
\Theta_\Delta(w,v)/ {\cal V} \epsilon^3$ where the Theta functions are
one if the first argument is within any of the eight tetrahedra surrounding 
the vertex $v$ and zero otherwise, and the volume of each tetrahedra is
given by $\epsilon^3{\cal V}$.
This expression is quite similar to the original proposal for a
(doubly densitized) Hamiltonian in the loop representation in terms of
the loop derivative \cite{Ga91}. If one particularizes this expression
to the expectation value of the Wilson net, one gets a very compact expression 
\cite{DiGaGrPu99},
\begin{equation}\label{nu}
H(N) E
\left(
\raisebox{-10mm}{\psfig{file=ver3.eps,height=20mm}},\kappa\right) = 
-{\kappa \over 3G}
\sum_{v\in s} N(v)  \nu_{(J_1J_2J_3)}
E\left(
\raisebox{-10mm}{\psfig{file=ver3.eps,height=20mm}},\kappa
\right), \nonumber
\end{equation}
where $\nu_{J_iJ_jJ_k}$ is a group factor. From this expression one can
derive the action of the Hamiltonian on a given Vassiliev invariant; it
turns out to produce an invariant of one order less.  
It is quite remarkable that
the action of the loop derivative in a space of diffeomorphism
invariant functions yields a finite well defined expression for the
constraint.  For intersections of valences higher than three the
action of the Hamiltonian ceases to be just a prefactor, but it still
can be written explicitly. One can also introduce a diffeomorphism
constraint,
\begin{equation}
C(\vec{N}) \Psi(s) = \sum_k
\lim_{\epsilon\rightarrow 0} \int d^3x \int_{e_k} dy^b
{(N^a(x)+N^a(y))\over 2}f_\epsilon(x,y) \Delta_{ab}(\pi_y^x)
\Psi(s).
\end{equation}
where $f_\epsilon(x,y)$ is a regularization of the Dirac delta.
Acting on Vassiliev invariants, one can explicitly check via a detailed
calculation \cite{DiGaGrPu99} that the constraint vanishes identically,
as one would expect since the wavefunctions are diffeomorphism invariant
\cite{foot0}.

As we see from equation (\ref{hgen}), the action of the Hamiltonian
constraint on a Vassiliev invariant produces a  prefactor that depends on the
location of the vertices times a group prefactor times a Vassiliev
invariant. The location of the vertex is determined by the the
intersection of the edges of the spin network.  The latter are
modified by the loop derivative, and as a consequence the loop
derivative acts on functions of the position of the vertices. The loop
derivative leaves the group factors unchanged. Therefore the action of
the Hamiltonian produces as a result a function that is not
diffeomorphism invariant but that is still loop
differentiable, allowing  one can compute the constraint algebra.
We call these states generically $\psi(s,M,\Omega)$ where $M$
is the function of the vertex and $\Omega$ the group factor. We can
think of these states as the action of an operator $\hat{O}(M,\Omega)$
on $\psi(s)$. An explicit calculation \cite{DiGaGrPu99} shows that,
\begin{equation}
C(\vec{N}) O(M,\Omega)\psi(s)
= O(N^a \partial_a M,\Omega)
\psi(s) + O(M,\Omega) C(\vec{N}) \psi(s).
\label{diffeoono}
\end{equation}
That is, the diffeomorphism Lie-drags the prefactor and therefore acts
geometrically. This ensures that the constraint algebra of
diffeomorphisms is correctly implemented in this space. It also shows that the
commutator of diffeomorphism and Hamiltonian is correct, that is, the
Hamiltonian transforms covariantly.

To study the consistency of the commutator of two Hamiltonians with
the classical Poisson relation $\{H(N),H(M)\}=C(q^{ab} V_a)$ where
$V_a = M\partial_a N-N\partial_a M$, one needs to promote to a quantum
operator the right-hand-side of the relation, which is proportional to
the product of a diffeomorphism and the doubly-contravariant spatial
metric. When one computes the right hand side, one finds that it vanishes
identically on spin network states. This, in fact, can be tracked down
to the vanishing of the double contravariant metric, which quantum
mechanically can be written as \cite{QSD3,DiGaGrPu99},
\begin{equation}
\hat{q}^{ab}(z)\psi(s)=\lim_{\delta\rightarrow 0}
\lim_{\epsilon\rightarrow 0} 
{ 8
 \over 9 G^2}  \sum_{v\in s} 
\int_{e_r} dy^a \int_{e_u} dw^b 
\epsilon^{pqr} \epsilon^{stu}
{\delta^6 \over \epsilon^6}
\Theta_\Delta(y,v) \Theta_{\Delta}(w,v) 
Q(e_p,e_q,e_s,e_t)
\psi(s)
\end{equation}
where the operator $Q$ can be written in terms of the holonomies along
edges incoming to the vertex and the volume operator and is finite for
any spin network. If one assumes that the regularizations $\delta$ and
$\epsilon$ are of the same order, the above expression is of order
$\epsilon^2$ (given by the two one dimensional integrals of $\Theta$
functions of size $\epsilon$) and therefore vanishes. If one computes
the doubly-covariant metric one finds that it diverges.  

In spite of the fact that the loop derivative acts on the prefactor
generated by the action of the Hamiltonian, when one computes the
successive action of two Hamiltonians a cancellation takes place 
\cite{DiGaGrPu99} and the  left hand side of the 
commutator equation vanishes  and therefore the algebra is
consistent. 

There is regularization ambiguity in these expressions. A clear
example of this is in the double contravariant metric where there are
two limits and one could choose to carefully ``tune'' them in order to
end with a non-vanishing expression. The price to pay is that the
non-vanishing expression depends on the background structures used in
the regularization. This is not surprising. In the spin network
representation we are in a manifold without a pre-determined
metric. The only information we have are the locations of
intersections and the orientations of the lines entering (not their
tangent vectors). This is insufficient information to construct a
symmetric tensor.  Therefore the expression for the metric was bound
to either be zero or background dependent. Similar considerations hold
for the covariant metric. A posteriori, the result we find via a
careful regularization is what one should have intuitively expected.

We therefore have a non-trivial, well defined quantization of 
canonical general relativity with the space of states given by the
Vassiliev invariants. The expressions of the constraints are
relatively simple, well defined and finite. Moreover, one can compute
the constraint algebra and it is consistent with the classical Poisson
algebra. Notice that the realization of the constraints is ``off
shell'' in the sense that we do not need to work with diffeomorphism
invariant states from the outset, and in fact this is sensible since
the Hamiltonian constraint does not map within such a space of
states. These points (the space of states chosen and the fact that we
have an infinitesimal generator of diffeomorphisms) distinguish our
construction from that of Thiemann which operated on diffeomorphism
invariant states. It has in common the fact that the Hamiltonian
commutes with itself. 

Should one worry about a theory of quantum gravity where the metric
appears to vanish? This will largely depend on how the semi-classical
limit is set up for the theory. As we argued above, the double
contravariant metric could not be anything else but vanishing in the
context of the spin network quantum theory. More meaningful physical
operators (like the length, the area and curvature invariants
\cite{foot}) are
non-vanishing and the volume operator would also be non-vanishing if
one included intersections beyond the trivalent ones. A correct
semi-classical limit could be built in terms of these and other
operators which are in no sense degenerate. Can one find solutions to
the Hamiltonian constraint? We can already construct several. If one
considers the framing independent Vassiliev invariants, one can check
that they are annihilated by the Hamiltonian constraint (in the
context of trivalent intersections) \cite{DiGaGrPu99}. What is
lacking if one compares with the construction of Thiemann is to have
an inner product that would allow us to characterize these and other
states as normalizable. Other, more non-trivial solutions (some of
them with a cosmological constant) are likely to be present, as is
hinted by the results involving Chern--Simons states in the loop
representation (\cite{BrGaPunpb}, see also \cite{GaGrPu98} for some
results in terms of spin networks).

Thiemann's approach has also been studied in $2+1$ dimensions
\cite{QSD4}, and appears to lead to a satisfactory quantization,
provided one chooses in an ad-hoc way an inner product that rules out
certain infinite dimensional set of solutions. In a forthcoming paper
we will discuss the quantization of $2+1$ dimensional gravity using an
approach that has elements in common with the one we pursue here, in
particular the requirement of loop differentiability of the states.
We will see that this requirement limits us (at least for
low valence intersections) to the correct solution space in a natural
way.

Having a family of consistent theories provides a context for
calculations that are of a more ``kinematical'' nature, like the
calculations of the entropy of black holes \cite{AsBaCoKr}. It also
provides a basis for calculations of semi-classical behavior that are
more dependent on the dynamics of the theory \cite{GaPu98}. It is
expected that the theory could be coupled to matter 
following the ideas of Thiemann \cite{QSD5}. Deciding if one of these
consistent theories is a physically realistic quantum theory of
gravity will have to wait until testable predictions
that involve the dynamics in a more elaborate way are worked out.

We wish to thank Abhay Ashtekar, Laurent Freidel, Don Marolf 
and Thomas Thiemann
 for comments and discussions.  This work was supported in part by the
 National Science Foundation under grants PHY-9423950, INT-9811610,
 PHY-9407194, research funds of the Pennsylvania State University and
 the Eberly Family research fund at PSU.  JP acknowledges support of
 the Alfred P. Sloan and John S.  Guggenheim foundations. We
 acknowledge support of PEDECIBA.


\begin{references}
\bibitem{foot0}
If one acts on framing-dependent states, with the chosen
regularization for the diffeomorphism constraint the constraint still
vanishes. One can choose other regularizations, less symmetric on the
dependence on the smearing vector, which do not vanish on
framing-dependent invariants.
\bibitem{foot} In the Vassiliev context, curvature invariants, like
$\int d^3x \sqrt{{\rm det q}} R_{ab}R^{ab}$ can be properly defined 
since the curvature has a natural action defined through the loop
derivative, which in turn is dual to the magnetic field constructed
with the Ashtekar connection. This leads
to the above invariant being related to the volume operator.
\end{references}
\end{document}